# The critical behaviour of Ising spins on 2D Regge lattices*

*Christian Holm*[1] and *Wolfhard Janke*[1,2]

[1] Institut für Theoretische Physik, Freie Universität Berlin
Arnimallee 14, 14195 Berlin, Germany

[2] Institut für Physik, Johannes Gutenberg-Universität Mainz
Staudinger Weg 7, 55099 Mainz, Germany

**Abstract**

We performed a high statistics simulation of Ising spins coupled to 2D quantum gravity on toroidal geometries. The tori were triangulated using the Regge calculus approach and contained up to $512^2$ vertices. We used a constant area ensemble with an added $R^2$ interaction term, employing the $dl/l$ measure. We find clear evidence that the critical exponents of the Ising phase transition are consistent with the static critical exponents and do not depend on the coupling strength of the $R^2$ interaction term. We definitively can exclude for this type of model a behaviour as predicted by Boulatov and Kazakov [Phys. Lett. **B186**, 379 (1987)] for Ising spins coupled to dynamically triangulated surfaces.

*Work supported in part by the EEC under contract No. ERBCHRX CT93043.

# 1 Introduction

With the advances made in string theory there was a rising interest in 2D quantum gravity, motivated by the fact, that a string moving in $N$ dimension is equivalent to $N$ scalar fields coupled to 2D quantum gravity [1]. For $N$ equal to zero we then deal with pure gravity. An important step forward was made when Kazakov [2] suggested a model of Ising spins living on the vertices of $\phi^3$ graphs that was solvable, and in its dual form, equivalent to Ising spins coupled to a dynamical triangulated surface (DTS). It turned out [3] that the set of critical exponents of the Ising transition is very different from the static Onsager critical exponents (see Table 1 below). Some time later Knizhnik *et al.* (KPZ) [4] found the same set of critical exponents with methods of conformal field theory for matter of central charge $c = 1/2$, which is supposed to be the continuum limit of Kazakov's model. The exponents have been confirmed in a variety of numerical studies [5–8] on $\phi^3$ graphs as well as with the dual method of dynamical triangulated surfaces.

One of the oldest methods to study general relativity numerically was suggested by Regge [9]. He uses a simplicial approximation to a manifold, where the underlying lattice has fixed coordination number, and takes the link lengths as gravitational degrees of freedom, whereas in the DTS method it is just the opposite, the coordination number varies and the simplices have fixed link lengths. The relation of Regge calculus to classical relativity has been extensively studied [10, 11], and its continuum limit is reasonably well understood. There has been up to now only one simulation [12] with rather low statistics of Ising spins coupled to gravity using the Regge approach, which suggests, that the critical exponents are the same as in the flat case. This is somewhat disturbing, because it is generally believed, that both approaches should describe the same model in the continuum limit. We found it therefore worthwhile to investigate the Regge approach again on larger lattices and with a better statistics, to confirm or disprove the results obtained in Ref. [12].

# 2 The model and simulation techniques

We simulated the gravitational interaction using the Regge calculus where the underlying manifold is discretized with a fixed triangulation and the link lengths are subject to variations. We used the usual transcription [13, 14] of continuum quantities like the metric $g$ and the scalar curvature $R$ into the Regge approach, namely

$$\int d^2x\sqrt{g(x)} \longrightarrow \sum_i A_i, \tag{1}$$

$$\int d^2x\sqrt{g(x)}R(x) \longrightarrow 2\sum_i \delta_i, \tag{2}$$

$$\int d^2x\sqrt{g(x)}R^2(x) \longrightarrow 4\sum_i \frac{\delta_i^2}{A_i}, \tag{3}$$

where $\delta_i$ is the deficit angle at the vertex $i$ defined as

$$\delta_i = 2\pi - \sum_{\text{all } t \text{ sharing } i} \theta_i(t), \tag{4}$$

and $\theta_i(t)$ is the dihedral angle associated with the triangle $t$. The area $A_i$ is taken to be the baricentric area associated with the site $i$,

$$A_i = \sum_{t \supset i} \frac{1}{3} A_t, \tag{5}$$

where $A_t$ denotes the area of the triangle $t$, which is one of several popular choices that are believed to be equivalent in the continuum limit. We simulated the partition function

$$Z = \sum_{\{s\}} \int D\mu(l) \exp\left(-I(l) - KE(l,s)\right) \tag{6}$$

where $I(l)$ is defined as the gravitational action

$$I(l) = \sum_i (\lambda A_i + a\frac{\delta_i^2}{A_i}), \tag{7}$$

and

$$E(l,s) = \frac{1}{2} \sum_{\text{edges } l_{ij}} A_{ij} (\frac{s_i - s_j}{l_{ij}})^2 \tag{8}$$

is the energy of Ising spins $s_i$, $s_i = \pm 1$, which are located at the vertices $i$ of the lattice. Here the volume $A_{ij}$ associated with a link $l_{ij}$ is defined as

$$A_{ij} = \sum_{\text{triangles } t \supset l_{ij}} \frac{1}{3} A_t. \tag{9}$$

In two dimensions the Einstein-Hilbert action (2) is in virtue of the Gauss-Bonnet theorem proportional to the Euler characteristic $\chi$, and therefore a topological invariant which does not contribute to the gravitational action (7). In higher than two dimensions a $R^2$ interaction term is sometimes added to guarantee the boundedness of the gravitational action from below. This is not necessary in two dimensions where we included such a term to probe its influence on the continuum limit.

The form of the path integral measure $D\mu(l)$ in (6) is already not clear in the continuum formalism. The most simple and most often used choice on the lattice is $D\mu(l) = \prod dl/l$, which we also adopted here. We simulated the gravitational action using the standard single-hit Metropolis update. In addition to the usual Metropolis procedure a change in link length is only accepted, if the links of a triangle fulfill the triangle inequality. As Ising update we used the single-cluster (Wolff) algorithm [15] which prevents the critical slowing down near the phase transition. Between measurements we performed $n = 2 \ldots 4$ Monte Carlo steps consisting of one lattice sweep to update the link lengths $l_{ij}$ followed by a single-cluster flip to update (a fraction of) the spins $s_i$. We checked in some cases that varying the relative frequency of link and spin updates does not change the results within error bars.

We simulated the partition function (6) on triangulated tori of size $N = L^2$ with fixed coordination number $q = 6$. This gives rise to $2N$ triangles and $3N$ link variables. The principal simulations were performed at $a = 0.001$ and the couplings $K = 1$ and $K = 1.025$ for $L = 6, 8, 10, 12, 16, 32, 64, 100, 128,$ 200, 256, and 512. Additional simulations were performed with $a = 0$ and 0.1 at $K = 1.025$, using lattices of size $L = 8, 16, 32, 64, 100, 128$ and 256. Because of the scale invariance of the measure we could rescale each link when proposing a link update such that the total area was kept fixed to its initial value $A = N$. The difference of the model defined by (6) and the Ising model on a static triangular lattice is that the spins are coupled by geometric weight factors $w_{ij} = A_{ij}/l_{ij}^2$ which can fluctuate around the static value $w_{ij}^0 = \sqrt{3}/6$, with a peak at $w_{ij} < w_{ij}^0$ and a long tail for large $w_{ij}$. With decreasing coupling $a$ the peak sharpens and its location shifts to small values of $w_{ij}$.

For each run we recorded the time series of the energy density $e = E/A$, the magnetization density $m = \sum_i A_i s_i / A$ and the Liouville field density $\varphi = \sum_i \ln A_i/A$. For each lattice we performed about 50000 measurements. From an analysis of the time series we found integrated autocorrelation times for the energy and the magnetization of about 1−4 (in units of measurements) for all lattice sizes. To obtain results for the various observables $\mathcal{O}$ at $K$ values in an interval around the simulation point $K_0$, we applied the reweighting method [16]. Since we recorded the time series this amounts to computing

$$\langle \mathcal{O} \rangle |_K = \frac{\langle \mathcal{O} e^{-\Delta K E} \rangle |_{K_0}}{\langle e^{-\Delta K E} \rangle |_{K_0}}, \tag{10}$$

with $\Delta K = K - K_0$. To obtain errors we devided each run into 20 blocks and computed standard Jackknife errors. At $a = 0.001$ where we had two simulations at different $K$ values, we combined the results according to their errors [17, 18].

From the time series we computed the Binder parameter [19],

$$U_L(K) = 1 - \frac{1}{3} \frac{\langle m^4 \rangle}{\langle m^2 \rangle^2}. \tag{11}$$

It is well known that the $U_L(K)$ curves for different $L$ cross around $(K_c, U^*)$ with slopes $\propto L^{1/\nu}$, apart from confluent corrections explaining small systematic deviations. This allows an almost unbiased estimate of the critical coupling $K_c$, the critical correlation length exponent $\nu$, and the renormalized charge $U^*$. The slopes can be conveniently calculated as

$$\frac{dU_L}{dK} = (1 - U_L) \left\{ \langle E \rangle - 2 \frac{\langle m^2 E \rangle}{\langle m^2 \rangle} + \frac{\langle m^4 E \rangle}{\langle m^4 \rangle} \right\}. \tag{12}$$

We further analyzed the (finite lattice) susceptibility,

$$\chi(K) = A(\langle m^2 \rangle - \langle |m| \rangle^2), \tag{13}$$

the susceptibility in the disordered phase,

$$\chi'(K) = A(\langle m^2 \rangle), \tag{14}$$

the specific heat,

$$C(K) = K^2 A(\langle e^2 \rangle - \langle e \rangle^2), \tag{15}$$

and studied the (finite lattice) magnetization at its point of inflection, $\langle|m|\rangle|_{\rm inf}$. The inflection point can be obtained from the maximum of $d\langle|m|\rangle/dK$, which can be calculated as

$$\frac{d\langle|m|\rangle}{dK} = \langle E\rangle\langle|m|\rangle - \langle E|m|\rangle. \tag{16}$$

Further useful quantities are the logarithmic derivatives

$$\frac{d\ln\langle|m|\rangle}{dK} = \langle E\rangle - \frac{\langle E|m|\rangle}{\langle|m|\rangle}, \tag{17}$$

and

$$\frac{d\ln\langle m^2\rangle}{dK} = \langle E\rangle - \frac{\langle Em^2\rangle}{\langle m^2\rangle}. \tag{18}$$

Another gravitational quantity of interest is the Liouville field $\varphi(x) = \ln\sqrt{g(x)}$. In the discretized version its lattice average reads as $\varphi = 1/A\sum_i \ln A_i$, and the associated lattice Liouville susceptibility is defined as

$$\chi_\varphi(L) = A(\langle\varphi^2\rangle - \langle\varphi\rangle^2). \tag{19}$$

# 3 Results

By applying reweighting techniques we first determined the maxima of $\chi$, $C$, $d\langle|m|\rangle/dK$, $d\ln\langle|m|\rangle/dK$, and $d\ln\langle m^2\rangle/dK$. The location of the maxima provided us with five sequences of pseudo-transition points $K_{\rm max}(L)$ for which the scaling variable $x = (K_{\rm max}(L) - K_c)L^{1/\nu}$ should be constant. Using this information we then have several possibilities to extract the critical exponent $\nu$ from (linear) least square fits of the finite-size scaling (FSS) Ansatz[1] $dU_L/dK \cong L^{1/\nu}f_0(x)$ or $d\ln\langle|m|^p\rangle/dK \cong L^{1/\nu}f_p(x)$ to the data at the various $K_{\rm max}(L)$. The fit range was chosen such that the goodness-of-fit parameter $Q$ was always above 0.1.

For the very extensive simulations at $a = 0.001$ all values are in good agreement with the Onsager value $\nu = 1$ at a 2 % level, giving rise to an average of $1/\nu = 1.00(1)$. For the other two couplings, $a = 0.1$ and $a = 0$, the data scatter a bit more but their averages $1/\nu = 0.98(1)(a = 0.1)$ and $1/\nu = 0.95(2)(a = 0)$ are still compatible with $\nu = 1$.

[1] By writing the scaling Ansatz as $\cong N^{1/D\nu}$ instead of $\cong L^{1/\nu}$ we trivially get the same estimates for $2/\nu D$. This would be directly comparable with DTS simulations where $D$ is possibly a non-trivial internal fractal dimension; cp. Table 1.

Assuming thus $\nu = 1$ we have next determined estimates for $K_c$ from the Binder parameter crossings and the scaling of the various $K_{\rm max}(L)$. The crossings $K^\times$ of the curves $U_L(K)$ with $L$ and $L'$ approach $K_c$ as

$$K^\times = K_c + \kappa/(b^{1/\nu} - 1), \tag{20}$$

where $b = L'/L$, $\kappa$ is a constant, and confluent corrections are neglected. This method, valid for large $b$, turned out to be the most precise one, leading for $a = 0.001$ to an estimate of the critical coupling $K_c = 1.02652(6)$. For $a = 0.1$ we obtain by the same method $K_c = 1.0292(1)$, and for $a = 0$ we find $K_c = 1.0230(2)$. Performing linear least square fits with $\nu = 1$ to the various pseudo-transition points $K_{\rm max}(L)$ gave consistent estimates of $K_c$ for all five sequences. As final value we quote their average $K_c = 1.02650(8)$, where the error is taken as the minimum among the five estimates. This is probably an overestimate, but since the five sequences are of course correlated a more refined optimization of the final estimate would require a careful analysis of cross-correlations which we have not performed here. For the other couplings $a = 0.1$ and $a = 0.0$ we obtain $K_c = 1.0301(2)$, and $K_c = 1.0243(4)$, respectively, again in reasonable agreement with the estimate from the Binder parameter crossings. Combining this information we use in further analyses

$$K_c = 1.0234 \pm 0.0002 \quad (a = 0.0), \tag{21}$$
$$K_c = 1.0265 \pm 0.0001 \quad (a = 0.001), \tag{22}$$
$$K_c = 1.0295 \pm 0.0001 \quad (a = 0.1). \tag{23}$$

In particular we can now extract $\nu$ also from the scaling of $dU/dK$ and $d\ln\langle|m|^p\rangle/dK$ at $K_c$, in good agreement with our previous results.

Another quantity of interest is the asymptotic limit $U^*$ of the Binder parameter at $K_c$. We first looked at the scaling of $U^\times = U(K^\times)$ for fixed $b = L'/L = 2$, where one can assume a scaling of the form

$$U^\times(K^\times) = U^* + cL^{-\omega}. \tag{24}$$

From a three-parameter fit we obtain $U^* = 0.616(9)$ and $\omega = 0.6(5)$ for $a = 0.001$. Another somewhat more stable way to obtain an estimate for $U^*$ is by fitting the values of $U_L$ at $K_c$ according to (24). For $a = 0.001$ we obtain from the three-parameter fit shown in Fig. 1 $U^* = 0.6117(14)$ and $\omega = 1.1(4)$. The uncertainty in $K_c$, however, implies a further error of $\Delta U^* \approx 0.0035$, so that we finally quote $U^* = 0.612(5)$. From three-parameter fits at $K_c$ we obtain for $a = 0.1$ $U^* = 0.615(6)$ and $\omega = 0.7(4)$. For $a = 0.0$ the errors turned out to be larger, leading to an estimate of $U^* = 0.609(33)$ and $\omega = 0.3(7)$.

Figure 1: Finite-size scaling of the Binder parameter values $U_L(K)$ for $a = 0.001$ and $K = 1.0265 \approx K_c$. The solid line is a three-parameter fit $U_L(K) = U^* + cL^{-\omega}$, yielding $U^* = 0.6117(14)$ and $\omega = 1.1(4)$.

These values are practicably indistinguishable from estimates for the regular square lattice, $U^* = 0.615(10)$ [20] and $U^* = 0.611(1)$ [21], or Poissonian random lattices, $U^* = 0.615(7)$ [22]. This gives further evidence that the critical behaviour of Ising spins on Regge lattices is governed by the standard (Onsager) universality class. We are not aware of any estimates for $U^*$ in the DTS approach.

To extract the critical exponent ratio $\gamma/\nu$ we used the scaling $\chi \cong L^{\gamma/\nu} f_3(x)$ at the previously discussed points of constant $x$, as well as the scaling of $\chi'$ at $K_c$. The quality of the fits for $\chi_{\max}$ can be inspected in Fig. 2. As final results we find $\gamma/\nu = 1.735(5)(a = 0.001)$, $\gamma/\nu = 1.728(3)(a = 0.1)$, and $\gamma/\nu = 1.717(6)(a = 0)$. The values for $\gamma/\nu$ for the different values of $a$ are compatible with each other, but are all slightly below the Onsager value of $\gamma/\nu = 1.75$. Due to their respective error range, however, they are still consistent with the flat space exponent ratio.

To extract the magnetical critical exponent ratio $\beta/\nu$ we used that $\langle|m|\rangle \cong L^{-\beta/\nu} f_4(x)$ at all constant $x$-values. Another method is too look at the scaling of $d\langle|m|\rangle/dK \cong L^{(1-\beta)/\nu} f_5(x)$. Because the errors on the different estimates turned out to vary over a large range we chose to compute error-weighted averages. Using our average values for $\nu$ we obtain the final estimates of $\beta/\nu = 0.127(3)$ $(a = 0.001)$, $\beta/\nu = 0.123(2)$ $(a = 0.1)$, and $\beta/\nu = 0.123(4)$ $(a = 0.0)$. Again we see little influence of the curvature square term, and the results are again in agreement with the Onsager result $\beta/\nu = 0.125$.

Having found so far overwhelming evidence for the Onsager universality class, we expect also the specific-heat exponent $\alpha$ to take on the Onsager value, namely $\alpha = 0$. In this case we expect a logarithmic divergence like

$$C(x, L) = A(x) + B(x) \ln L. \tag{25}$$

Indeed the data at the different fixed values of $x$ can all be fitted nicely with this Ansatz. In particular, for a fit of $C_{\max}$ with 11 data points at $a = 0.001$ we obtain $A = 0.17(1)$, $B = 0.369(5)$, with $Q = 0.80$. Of course, we also did an unbiased 3-parameter fit of the form

$$C(x, L) = A(x) + B(x) L^{\alpha/\nu}, \tag{26}$$

which gave us in the case of the fit of $C_{\max}$ with $a = 0.001$ and 8 data points $A = 7.9(3.8)$, $B = -7.9(3.7)$, and $\alpha/\nu = -0.06(5)$, with a quality $Q = 0.12$, compare Fig. 3. For comparison, we included also the best fit with Ansatz (25), and a linear least-square fit with the DTS prediction $\alpha/\nu = -2/3$. For the other two $a$-values we get $A = 7.4(8.1)$, $B = -7.4(7.8)$, and $\alpha/\nu = -0.07(9)$, with a quality $Q = 0.08$ for $a = 0.1$, and $A = 7(12)$, $B = -7(12)$, and $\alpha/\nu = -0.06(13)$, with a quality $Q = 0.32$ for $a = 0.0$. Similar fits of $C$ at the other $K_{\max}(L)$ sequences as well as at $K_c$ gave consistent results. Another method, which in previous studies [23] gave improved estimates over (26), is to fit the energy density according to $e(L) = a + BL^{(\alpha-1)/\nu}$. Three-parameter fits at $K_c$ gave $(\alpha-1)/\nu = -0.98(5)(a = 0.001)$, $(\alpha-1)/\nu = -0.98(4)(a = 0.1)$, and $(\alpha-1)/\nu = -1.06(4)(a = 0.0)$. Overall we can conclude that also for $\alpha/\nu$ our data is fully consistent with the static Onsager value of zero.

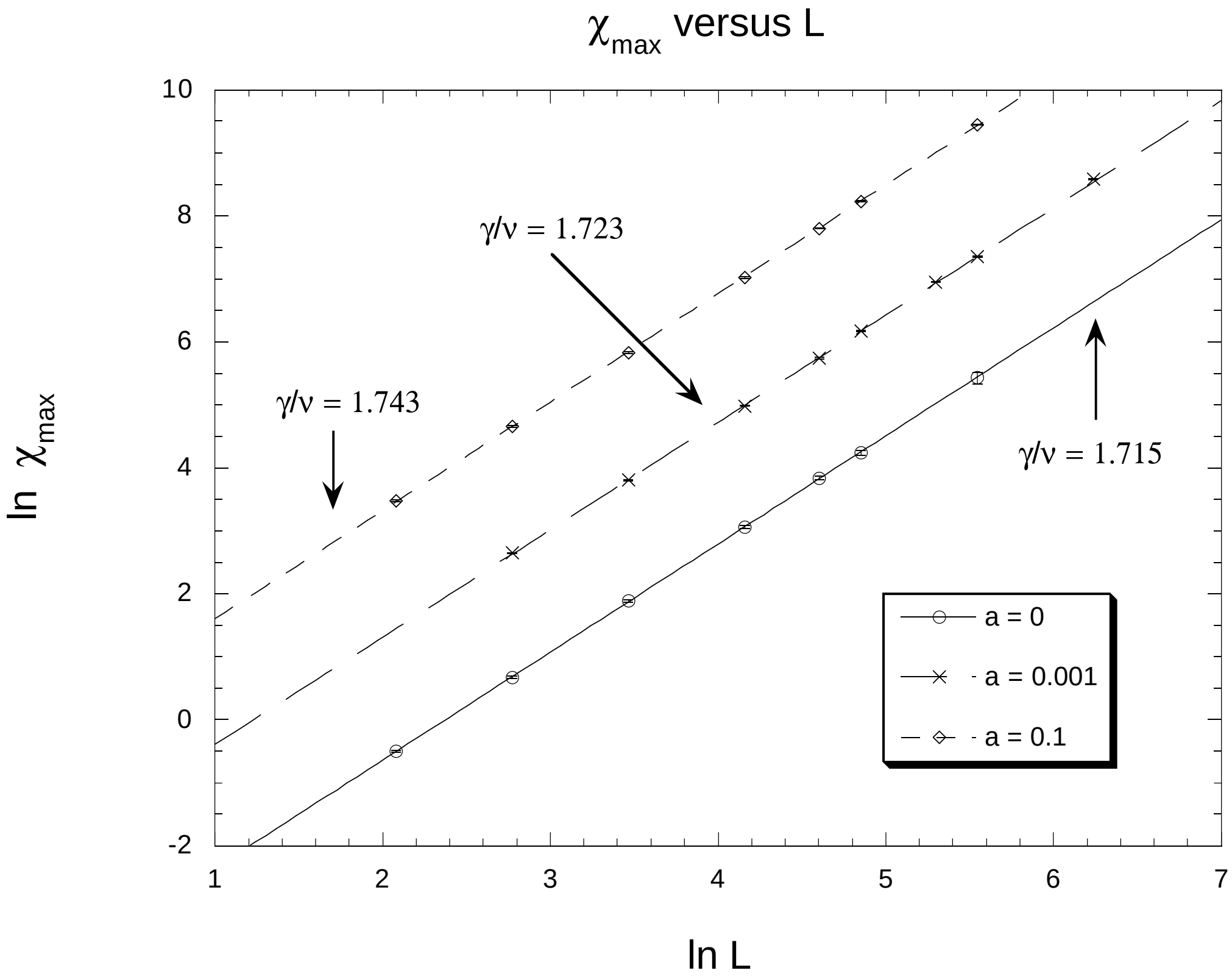

Figure 2: Double logarithmic finite-size scaling plot of the susceptibility maxima $\chi_{\max}$ for $a = 0.0, 0.001$, and $0.1$. To disentangle the curves we added an offset of $-2$ (2) to the data for $a = 0.0$ ($a = 0.1$). The slopes are in all three cases compatible with the Onsager value $\gamma/\nu = 1.75$ for regular static lattices.

We found in all FSS analyses that the added $R^2$ interaction term did not affect the FSS behaviour. To get a final estimate we therefore computed a weighted average of the three simulations with different coupling $a$. The results, put into comparison with the KPZ predictions, can be found in Table 1. We used the usual scaling relations $\eta = 2 - \gamma/\nu$ and $\delta = 1 + \gamma/\beta$, as well as our measured value for $\nu$ to convert the exponent ratios.

Let us finally consider the critical behaviour of the Liouville field $\varphi$ which is the interesting variable in the gravitational sector. The FSS prediction is $\chi_\varphi(L) \cong L^{(2-\eta_\varphi)} f_6(x)$ [12]. Using the 7 largest values for $L$ in our simulation for $a = 0.001$ at $K_c$ a linear-least-square fit yields $(2-\eta_\varphi) = 0.047(28)$, with $Q = 0.15$, and for $a = 0.1$ we get with 6 data points $(2-\eta_\varphi) = 0.040(13)$, with $Q = 0.79$. For $a = 0$, on the other hand, we find with 7 data points $(2-\eta_\varphi) = 2.01(6)$, with $Q = 0.19$. The observed pronounced dependence of $\eta_\varphi$ on $a$ is plausible because a positive $a$ term tends to suppress curvature fluctuations and stabilizes in this way also the area fluctuations. Therefore one should only consider the case with $a = 0$, which gives an observed value of $\eta_\varphi \approx 0$ that is consistent with a massless free field behaviour of $\varphi$.

# 4 Concluding remarks

We have performed a fairly detailed high statistic study of the Ising model coupled to quantum gravity via the Regge calculus approach. Using the path integral measure $\prod dl/l$ we have found that an included $R^2$ interaction term showed no effect on the nature of the phase transition, and that the critical exponents of the Ising transition still agree with the Onsager exponents for regular static lattices. If one believes in KPZ scaling then Regge calculus can survive as a tool to probe quantum gravity only, if one can modify it to reproduce the KPZ results. The two most likely alterations would be to implement a different coupling of the Ising spins to gravity, or to use a different, maybe more physically motivated, measure. In fact, exploratory simulations [24] with a different measure seem to imply that the choice of measure plays a more dominant role than has been thought so far.

Figure 3: Finite-size scaling of the specific-heat maxima for $a = 0.001$. Also shown are a logarithmic fit $C_{\rm max} = A + B \ln L$, a power-law fit $C_{\rm max} = A + B L^{\alpha/\nu}$, as well as a constrained power-law fit assuming the DTS prediction $\alpha/\nu = -2/3$.

## Acknowledgement

W.J. thanks D. Johnston for helpful discussions and the DFG for a Heisenberg fellowship. C.H. thanks H. Hamber for discussions and a sample code of their simulations. The numerical simulations were performed on the North German Vector Cluster (NVV) under grant bvpf01. We thank all institutions for their generous support.

| | $\alpha$ | $\beta$ | $\gamma$ | $\delta$ | $\eta$ | $\nu$ |
|---|---|---|---|---|---|---|
| DTS | $-1$ | 0.5 | 2 | 5 | $2/3^*$ | $1.5^*$ |
| Onsager | 0 | 0.125 | 1.75 | 15 | 0.25 | 1 |
| this MC study | $\approx 0$ | 0.126(2) | 1.75(2) | 14.9(3) | 0.272(3) | 1.01(1) |

Table 1: Comparison of our Monte Carlo results with the exact results for the Ising model on static lattices (Onsager) and the results of Boulatov and Kazakov [3] for the Ising model on dynamical $\phi^3$ graphs (DTS). The values marked with a star were computed from hyperscaling relations with $D = 2$, thereby neglecting possible scaling effects due to the internal fractal dimension in the DTS approach.